\documentclass[prl,a4paper,twocolumn,showpacs,preprintnumbers,amsmath,amssymb,floatfix,amstex]{revtex4}
\newcommand{\ket}[1]{|\,#1\,\rangle}        		%
\newcommand{\nx}[1]{\mathbf{#1}}	    		%
\newcommand{\uxu}{U\otimes U^{\dagger}}			%
\newcommand{\FP}{{\cal L}}				%
\newcommand{\FPe}{{\cal L}_{\epsilon}}			%
\newcommand{\cDe}{{\cal D}_{\epsilon}}			
\newcommand{\De}{\text{\textbf{\textsf{D}}}_{\epsilon}}	
\newcommand{\cpq}{c_{\epsilon}(p,q)}			%
\newcommand{\clm}{\tilde{c_{\epsilon}}(\lambda,\mu)}	%
\newcommand{\qL}{\text{\textbf{\textsf{L}}}}		%
\newcommand{\doll}{\text{\textbf{\textsf{L}}}_{\epsilon}}%
\newfont{\Bb}{msbm10}					%
\newcommand{\Ldos}{\mbox{\Bb\symbol{76}}^2}		%
\newcommand{\oT}{\hat{T}}				%
\newcommand{\tr}{\mbox{tr}}				%
\usepackage{graphicx,graphics,wrapfig,rotating}	
\usepackage{dcolumn}				
\usepackage{bm}					

\begin{document}


\title{ Classical decays in decoherent quantum maps}

\author{Ignacio Garc\'\i a-Mata}
\email{garciama@tandar.cnea.gov.ar}
\author{Marcos Saraceno}
\email{saraceno@tandar.cnea.gov.ar}
\author{Mar\'\i a Elena Spina}%
\email{spina@tandar.cnea.gov.ar}
 \affiliation{%
Dto. de F\'\i sica, Comisi\'on Nacional de Energ\'\i a At\'omica.
Libertador 8250 (1429), Buenos Aires, Argentina.
}%

\date{\today}

\begin{abstract}
We study the asymptotic long-time behavior of open quantum maps and relate the decays to the eigenvalues
of a coarse-grained superoperator. In specific ranges of coarse graining, and for chaotic maps, these
decay rates are given by the Ruelle-Pollicott resonances of the classical map. 
\end{abstract}

\maketitle
The study of the emergence of classical features in the quantum
behavior of hamiltonian systems has centered on two areas: on the one
hand, the statistical
fluctuation properties of the quantum spectrum were the first quantities to
reveal universality 
classes related to classical - chaotic or integrable - behavior \cite{bohi}.
On the other hand, the study of time
dependent quantities has also provided quantum signatures of the classical world.
The growth of Von Neuman entropy in decohering quantum
systems \cite{zurek,monte}  showed striking differences
separating chaotic and regular systems and pointed to the relevance of Lyapunov exponents
- more precisely the Kolmogorov-Sinai entropy - for this quantity's time behavior. Recently 
there has been a strong interest in the
study of the Loschmidt echo \cite{losch} - specially in the context of
fidelity decay in quantum maps and
algorithms - which displayed essentially the same characteristics.
In this letter we report the emergence and the relevance in quantum decohering maps of yet other 
quantities of a classical nature - the Ruelle-Pollicott resonances \cite{ruel,ruel2} - 
and show that they rule the asymptotic decay of time dependent quantities.

In \cite{bian} a simple model for decoherence in quantum 
maps was introduced based on the Kraus representation 
of superoperators \cite{Kraus}, with a straightforward 
interpretation as quantum coarse-graining in phase space. 
The model was employed to study the entropy evolution 
and to confirm the prediction \cite{zurek} that for certain 
regimes the rate of growth is independent of the coarse 
graining and is given by the Lyapunov exponent of the 
classical motion. This is a short time regime, of the  
order of the Ehrenfest time, after which relaxation towards 
the uniform density occurs.  
In this letter we study this asymptotic relaxation 
regime for quantum maps and show that it is ruled by the 
 eigenvalues of a similar coarse-graining superoperator. 
These eigenvalues can be effectively calculated  
and they determine 
- in certain well defined ranges - classical decay rates independent of the 
coarse graining. These decay rates are given, in the semiclassical 
limit of chaotic systems, by the Ruelle-Pollicott resonances. 
We show in particular that, beyond the Ehrenfest time, 
 these resonances determine the  
asymptotic behavior of the linear entropy and of the Loschmidt 
echo. 

The unitary evolution of the density matrix for a quantum system is
given by $ \rho_{n+1}=U\rho_{n} U^{\dagger}$
where $U$  is a unitary evolution operator that quantizes a classical map. For
simplicity we consider area preserving maps on a phase space with periodic boundary
conditions, {\it i.e.} a torus (normalized to unit area). In this case the Hilbert 
space has a finite dimension $ N $  related to   $\hbar$ by $ 2\pi\hbar N= 1 $.
Position and momentum eigenstates are respectively:
\begin{equation}
\begin{array}{lr}
\ket{q_i}&=\ket{i/N} \\
\ket{p_j}&=\ket{j/N}
\end{array}
\ \ i,j \in [0,N-1].
\end{equation}
They form an $N\times N$ grid that constitutes the quantum phase space for these maps
and their overlap is the discrete Fourier transform (DFT) of dimension $N$. Translations on this
grid can be implemented by unitary operators $\oT(p,q)$ , easily constructed from cyclic
shifts and the DFT \cite{Schwinger}. They carry the integer labels $p,q$ and they form a complete
unitary operator basis in Hilbert space. 
The linear action  of $U$ on the vector space of density matrices defines a
{\em superoperator\/} $\qL=\uxu$, such that
\begin{equation}
\rho_{n+1}=\qL \rho_{n}
\end{equation}
which is still unitary, of dimension $N^2\times N^2$. The eigenvalues of
$\qL$ are $ e^{i(E_i-E_j)} $, where  $E_i $ are the Floquet eigenvalues of the map.

The {\em quantum\/} coarse graining is implemented by a Kraus
superoperator
\begin{equation}
\De=\sum_{p,q=0}^{N-1} \cpq\,\oT(p,q)\otimes\oT^{\dagger}(p,q),
\end{equation}
 where $\cpq$ is a smooth positive periodic function on the torus
to be 
defined below. In order
for $\De$ to be trace preserving the condition $\sum_{p,q} \cpq =1$
must be satisfied. 
Inasmuch as  $\cpq$ is a narrow gaussian-like function, $\De$ implements an incoherent sum
of slightly displaced density matrices over a 
phase space region of order $\epsilon$, this being a measure of the amount of coarse graining.
To avoid an overall drift we further assume that $\cpq$ is an even function of its arguments.
The spectral properties of $\De$ are easily calculated. $\De$ is 
hermitian and using the fact that 
$ \oT(p,q)\oT(\lambda,\mu)\oT^{\dagger}(p,q)= 
\oT(\lambda,\mu)e^{i \frac{2\pi}{N}(\lambda q - \mu p) }$ 
its spectral properties are given by
\begin{equation}
\De \oT(\lambda,\mu)=\clm\oT(\lambda,\mu),
\end{equation}
where $\clm=\sum_{p,q}\cpq e^{i \frac{2\pi}{N}(\lambda q-\mu p)}$ is the DFT of $\cpq$. 
It is irrelevant whether we specify the diffusion by $\cpq$ or $\clm$. We found
more convenient to define the latter as
\begin{equation}
\clm=e^{-\frac{1}{2}(\frac{\epsilon
N}{\pi})^2(\sin^2[\pi\lambda/N]+\sin^2[\pi\mu/N])},
\end{equation}
whose Fourier transform is a smooth periodic gaussian-like coarse graining
kernel, of width $\epsilon/2\pi$.
The coarse-grained dynamics of the quantum map will be given by:
\begin{equation}
\rho_{n+1}=\De \qL \rho_{n}=\doll\rho_{n}.
\end{equation}
For any finite value of $\epsilon$ , $\doll$ is a convex sum of unitary matrices
and is therefore a {\it contracting } map whose spectrum  is contained in the
unit circle, and has a non degenerate eigenvalue 
equal to 1, corresponding to the uniform eigenvector $\oT(0,0)$, and $(N^2 -1)$
eigenvalues of modulus smaller 
than 1. Therefore, the time evolution of the
density matrix can be 
decomposed as a sum of exponentially decaying modes plus a constant. In particular, the
asymptotic decay towards the uniform density will be given by the eigenvalue closer to the 
unit circle. 
Although this is strictly true for any value of $N$ and $\epsilon > 0$, the behavior of
the eigenvalues in  the limits $ N \rightarrow\infty$ and $\epsilon \rightarrow 0$ is
strongly dependent on the classical features -chaotic or regular - of the map in question.

Before studying these limits we review the equivalent procedure for classical maps:
densities in phase space evolve with the Frobenius-Perron (F-P)
operator $\FP$ arising from the Liouville equation. For a classical map, the
general form of $\FP$ is $\FP(\nx{x},\nx{y})=\delta[\nx{y}-f(\nx{x})] $
where $\nx{x}=(q,p)$, $\nx{y}=(q',p')$. 
As it stands, the operator $\FP$ is unitary on $\Ldos$, the space of square
integrable functions on phase space. When it is convoluted with a narrow coarse
graining operator $\cDe$, $\epsilon$ being the coarse graining parameter,
we obtain 
\begin{equation}
\FPe=\cDe\circ\FP.
\end{equation}
$\cDe$ has the role of damping the high frequency components of phase space distributions 
in $\Ldos$ thus producing an {\em effective truncation\/} of $\FP$ and
drastically changing its spectrum. In fact now $\FPe$ is compact and its spectrum consists of
isolated resonances contained in the unit circle\cite{blank,nonn}. The behavior of these 
resonances as 
$\epsilon \rightarrow 0$ is radically different for chaotic or regular maps.
If the map is ergodic and mixing, $\FPe$ has 
an isolated non-degenerate eigenvalue equal to 1, corresponding to the
uniform density, and isolated eigenvalues with finite
multiplicity, and modulus strictly smaller than 1.
In the singular limit $\epsilon\rightarrow 0$, these eigenvalues are the
Ruelle resonances \cite{blank,nonn} and they determine the exponential 
decay of correlation functions in phase space \cite{ruel,ruel2}. 
An equivalent procedure to uncover these resonances consists in 
introducing a basis of
functions ordered by resolution  in phase space, then constructing 
$\FP$ in this basis 
and subsequently truncating to a finite dimension $M$, 
resulting in a matrix $\FP^{^{(M)}}$, whose
largest eigenvalues stay fixed as $M\rightarrow\infty$ (see
\cite{susy,webe,hase,khod},
\cite{fish} for a review and \cite{haak} for a quantum application).

It was proved recently, under quite broad conditions \cite{nonn} (Theorem 1), that for smooth
 maps on the torus and for any
fixed $\epsilon$ the spectrum of the coarse grained quantum propagator
$\doll $ converges to that of $\FPe$ in the semiclassical limit $N\to\infty$ . This in turn implies that 
at finite $\epsilon$
the quantum decay rates are given by the  corresponding classical eigenvalues. We study
below the issue of letting $\epsilon \to 0$.

We first compare these spectra for the perturbed Arnol\'d cat map
(and its quantization \cite{hannay})
\begin{equation}
		\label{pcat}
\begin{array}{rl}
p'=&p+q\ -2 \pi k\sin [2\pi q] \\
q'=&q+p'+2 \pi k\sin [2\pi p'] .
\end{array} \mbox{(mod 1)},
\end{equation}
The classical Arnold cat map is ergodic and mixing, has uniform hyperbolicity and  its Lyapuonov 
coefficient is $\chi=(1/2)(3+\sqrt{5})$. The perturbation is added to avoid
the non-generic behavior of the quantized version and $\chi$ is not changed substantially by it.
On the contrary, the Ruelle resonances are very sensitive to the perturbation.
\begin{figure}[h]
\begin{center}
\includegraphics*[width=4.5cm,angle=-90]{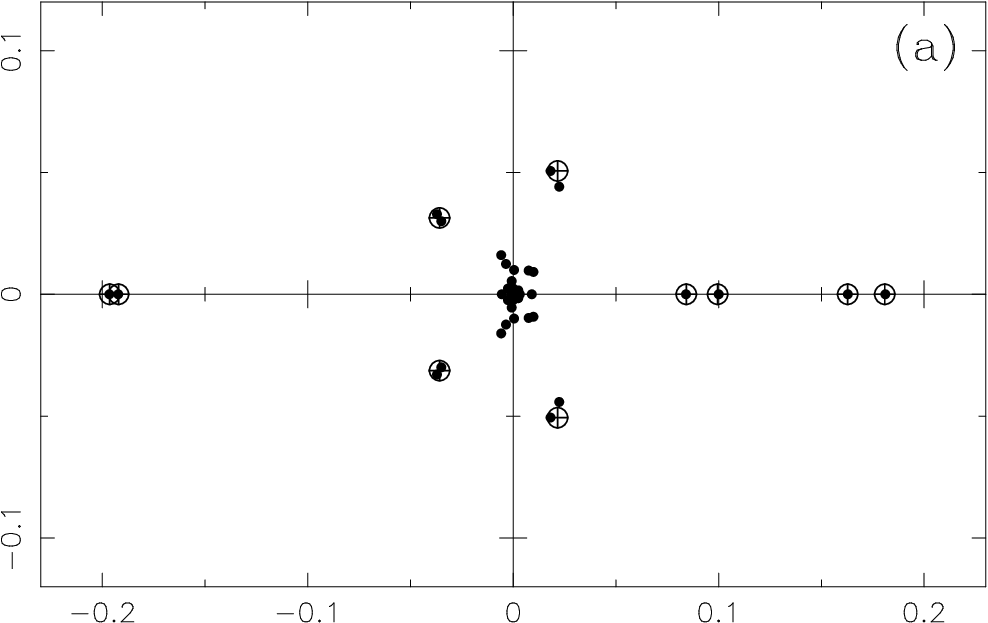}
\vspace{0.2cm}
\includegraphics*[width=4.5cm,angle=-90]{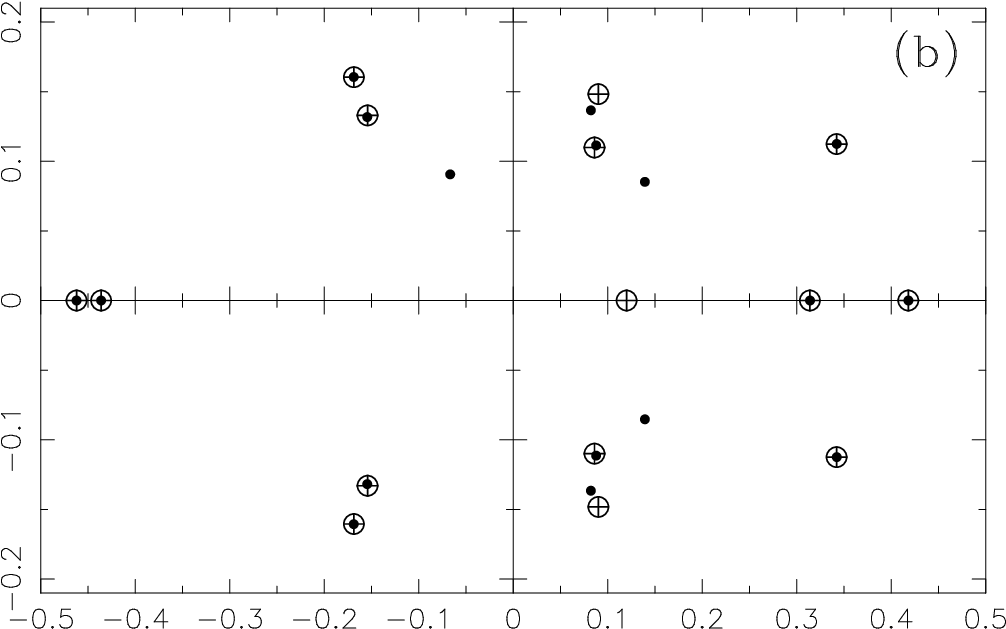}
\caption{%
Comparison of the leading eigenvalues of the quantum $\doll$ and
classical $\FPe$ coarse-grained propagators for the
perturbed Arnold cat map, with different coarse graining and perturbation
(a) $k=0.01$, $\epsilon=0.4$; (b) $k=0.02$, $\epsilon=0.314$.
($\bullet\equiv$ classical, grid of $50 \times 50$ ; $\oplus\equiv$
quantum, $N=150$). 
\label{spect} }
\end{center}
\end{figure}
In Fig \ref{spect} we compare the eigenvalues of $\FPe$ and $\doll$. The main difficulty 
in both calculations is the large dimensions of the matrices involved.
For the classical calculation we took the coarse graining operator $\cDe$
to be a periodic gaussian of width
$\epsilon/(2\pi)$. The spectrum of $\FPe$ was obtained by projecting eq.
(\ref{pcat}) on a grid of
$50\times50$ sites and diagonalizing. The procedure is stable for the leading eigenvalues 
as long as $\epsilon$ is much larger than the grid size \cite{blank}. The spectrum of $\doll$ was
obtained by a variational method similar to that proposed in \cite{agam}. A basis of 
right and left trial functions, which are smooth on the unstable and stable manifold
respectively, are used to construct a subspace that contains the leading eigenspace.
 The advantage of this method is that the resulting generalized eigenvalue problem has a
 small dimension as the basis contains information about the dynamics. 
Details about the computation of the spectrum of $\doll$ with this method
will be reported in a future article \cite{garma}.
 Good agreement for the leading eigenvalues of both spectra was
found in all the cases we computed, provided  $N$ was large enough,
in agreement with \cite{nonn}.
 In order to obtain the Ruelle resonances we now have to take the limit
$\epsilon\rightarrow 0$. Clearly this limit cannot be taken at constant $N$
because for finite matrices unitarity is eventually recovered and all eigenvalues go back to the unit 
circle. 

\begin{figure}[h]
\begin{center}
\includegraphics*[width=6.5cm,angle=-90]{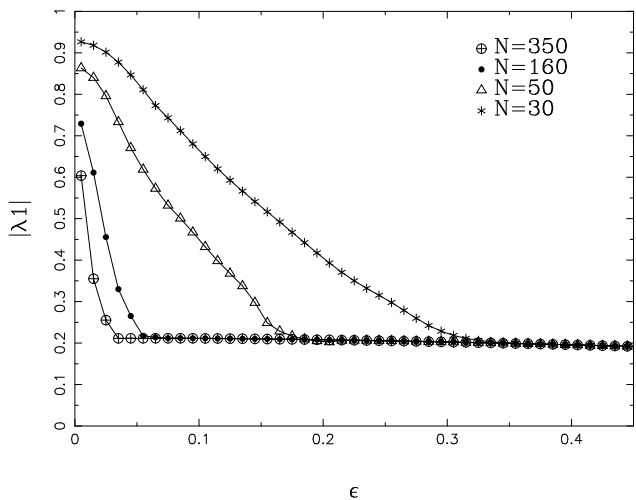}%
\vspace{-0.25cm}
\caption{%
The leading eigenvalue $|\lambda_1|$ as a function of  $\epsilon$,
for different values of $N$. There is a range, growing with $N$  where the eigenvalue is
independent of $\epsilon$ so that it can be extrapolated to
$\epsilon\rightarrow 0$. $k=0.01$.
\label{compare}}
\end{center}
\end{figure}

In Fig.~\ref{compare} we study this process as a function of $N$ and $\epsilon$ for the leading
eigenvalue $\lambda_1$. We observe that for each $N$ there is a range of values of
$\epsilon$ for which the eigenvalue is independent of  $\epsilon$ and for which
the limit   $\epsilon\rightarrow 0$  can be extrapolated safely, thus defining a property
of the classical map, independent of $N$ and $\epsilon$. A similar situation occurs for the
other eigenvalues, but the safe range becomes smaller as the distance to the unit circle increases.
We conclude therefore that the leading portion of the spectrum of $\doll$ for large $N$ is of a 
purely classical 
nature and provides the rates of asymptotic decay of quantum time dependent quantities. 
For chaotic maps these classical rates are further identified with the spectrum of 
Ruelle resonances. \cite{note}

\begin{figure}[h]
\begin{center}

\hspace{-0.5cm}
\includegraphics*[width=6cm,angle=-90]{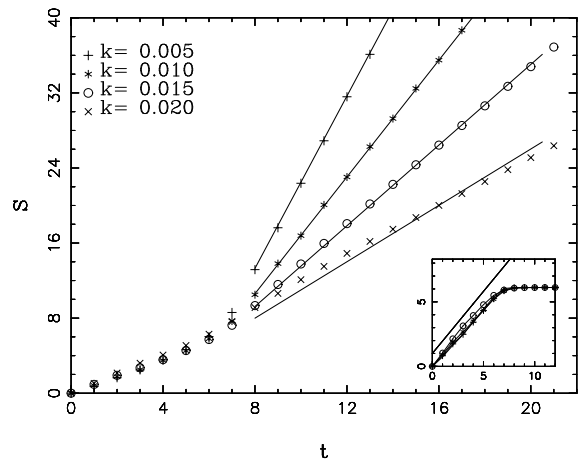}
\vspace{-0.25cm}
\caption{%
Linear entropy $S_n$ for the perturbed Arnold
cat for several values of $k$. The initial value is a coherent pure state
with the uniform density subtracted. $N=450,\ \epsilon=0.05$.
Notice the two linear regimes: the short time Lyapunov (inset) and the asymptotic Ruelle.
Solid lines are computed from the leading eigenvalue of $\doll$.
\label{entro}}
\end{center}
\end{figure}

To test this prediction consider first the evolution of the 
linear entropy $S_n=-\ln \tr[\rho_n^2]$ . It was shown in \cite{zurek,bian} that
for short times the entropy growth is linear with a slope  independent of
$\epsilon$ and determined by the Lyapunov exponent of the map.
This is shown in the inset of Fig. \ref{entro}, where  $S_n$ is plotted as
a function of $n$ for  $N=450$, $\epsilon =0.05$ and various values of the
perturbation.   As expected for an ergodic map, for longer times the state
relaxes to the uniform distribution and $S$ converges to 
$S_{\infty}=-\ln(1/N) $. To uncover the rate of relaxation we subtract the uniform
density from the initial state and obtain, 
using the long time decomposition of $\rho_n$ into decaying modes,
 $  S_n = - 2 \ln |\lambda_1| n $. Again a linear growth is obtained but this time
 with a slope dependent on the leading Ruelle resonance. This feature is clearly seen in
Fig. \ref{entro} where we plot $S_n$ for different values of $k$. 

\begin{figure}[h]
\begin{center}

\hspace{-0.5cm}
\includegraphics*[width=5.5cm,angle=-90]{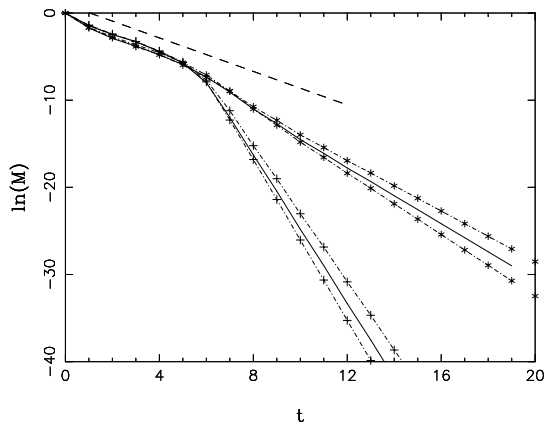}

\vspace{-0.25cm}
\caption{%
Decay of the Loschmidt echo beyond the Ehrenfest time. Two 
values of $k$,($+\equiv k=0.005$, $*\equiv k=0.017$ perturbed by
$\Delta k=0.002$ are plotted.$N=450,\ \epsilon=0.1$. The full line
shows the echo while the dotted lines give the perturbed and unperturbed
entropies $ -S_n$ and $ -S^\prime_n$.
\label{echo}}
\end{center}
\end{figure}

A very similar pattern emerges for the decay of the so called
Loschmidt echo, or fidelity \cite{losch}. This is defined as
$ M_n=\tr(\rho_n \rho^\prime_n) $ where $\rho_n$ and $\rho^\prime_n$
are states evolved with slightly different maps from an
initial state $\rho_0$. This quantity, originally considered
by Peres \cite{peres} as a measure of sensitivity to perturbations, has aquired renewed importance
as a measure of environment independent decay of spin echoes \cite{losch} and of the accuracy of 
implementation
of quantum algorithms. It is usually considered only in the case of strictly unitary
evolution where it equals $M_n(0) = \tr [\qL^n \rho_0 ~{\qL^\prime}^n \rho_0]$. For (slightly) open systems
it is natural to replace this quantity by $M_n(\epsilon) = \tr [\doll^n \rho_0 ~{\doll^\prime}^n \rho_0]$ 
and then consider the limit $\epsilon\to 0$ to extract features that are independent of the 
coarse-graining. The same considerations regarding the 
non-commutativity of the $ N $ and $\epsilon$ limits apply here
and in the ``safe" regime of $N$ and $\epsilon$ the asymptotic decay of $M_n$ is again ruled by the
Ruelle resonance. In the classical case, a similar result for the echo was obtained recently \cite{casat}.
Our discussion shows that the quantum decay must follow the classical one.
In fact the echo is closely related to the linear entropy discussed above. 
Schwartz inequality $ \tr(\rho_n \rho^\prime_n) \leq \sqrt{ \tr(\rho_n^2) \tr (\rho^{\prime 2}_n) }$ leads to
$\ln M_n\leq -(1/2)(S_n +S^\prime_n)$. Again the uniform density is subtracted from the initial state
to reveal the exponential relaxation regime. In the asymptotic regime the inequality becomes an 
equality and we find that beyond the Ehrenfest time the decay proceeds as
\begin{equation}
\ln M_n= -(1/2)(\ln|\lambda_1| +\ln|\lambda^\prime_1|) n.
\end{equation}
This is illustrated in  Fig. \ref{echo}, where it is clearly seen
that, after the initial Lyapunov regime the echo decays with the average of the two linear entropies. 

Our spectral analysis shows that for any observable whose quantum evolution depends on the operator $\qL$, 
and this is not restricted to maps only, the best procedure to extract its classical behavior is to
replace $\qL$ by $\doll$ and let $\epsilon \to 0$  and $N\to\infty $ (meaning $\hbar\to 0$ ) in such 
a way as to remain in the ``safe'' region where unitarity is not recovered. 
The convergence of the spectra of $\FPe$ and $\doll$, proven under very specific conditions 
in \cite{blank,nonn}, 
and demonstrated here in numerical experiments, is expected to be much more general, and in practice
provides a very powerful means of extracting classical behavior from quantum maps\footnote{Lately work
on the relation between classical and quantum trace formulas for maps has come to our attention 
\cite{braun}.}. 
Whether different models of decoherence, based
on physical mechanisms in small quantum devices, lead to the same results remains to
be explored  but, at least from the numerical point of view, are also subject to this spectral analysis
and can reveal other modes of decay not related to classical properties.

We are grateful to S. Nonnenmacher for communicating many of his results prior
to publication and to J.P.Paz for illuminating discussions.
Partial support is due to grants from ANPCYT, CONICET and ECOS-Secyt.

%

\begin{thebibliography}{99}
\bibitem{note} The situation is quite different for regular maps, because the classical
spectrum now consists of many eigenvalues converging to the unit circle as $\epsilon\to 0$.
Thus the decay is not exponential but power-law.
\end{thebibliography}
\end{document}